\documentclass[aps,pra,onecolumn,amsmath,amssymb,showpacs,superscriptaddress]{revtex4-2}

\usepackage{amsmath}
\usepackage{braket}
\usepackage{graphicx}% Include figure files
\usepackage{caption}
\usepackage{xcolor}
\usepackage{soul}
\usepackage{ulem}
\usepackage{cancel}
\usepackage{comment}
\usepackage[colorlinks=true, linkcolor=blue, citecolor=blue, urlcolor=blue]{hyperref}
\usepackage{booktabs}
\usepackage{upgreek}
\usepackage{bm}

\begin{document}

\title[]{Wavelength-adaptive spin-orbit orbital angular momentum management in three-wave mixing}

\author{Kiki Dekkers,$^{1,\dagger}$ Mwezi Koni,$^{2,\dagger}$ Vagharshak Hakobyan,$^{3}$ Sachleen Singh,$^{2}$ Jonathan Leach,$^{1}$ Etienne Brasselet,$^{3}$ Isaac Nape$^{2}$ and Andrew Forbes$^{2,*}$ \\
\vspace{5pt}
\small $^{1}$School of Engineering and Physical Sciences, Heriot-Watt University, Edinburgh, UK \\
\small $^{2}$School of Physics, University of Witwatersrand, Johannesburg, South Africa\\
\small $^{3}$University of Bordeaux, CNRS, Laboratoire Ondes et Mati{\`e}re d'Aquitaine, Talence, France \\
\vspace{5pt}
\small $^{\dag}$ These authors contributed equally to the work\\
\small $^{*}$ andrew.forbes@wits.ac.za\\
}

\begin{abstract}
    Here we propose the use of an adjustable liquid crystal spin-orbit device to shape bi-colour structured light to create bimodal states. We demonstrate the proof-of-principle for two individual wavelengths in a nonlinear optics framework.
    The spin-orbit device has an inhomogeneous optical axis orientation and birefringence, allowing it to modulate two wavelengths of light with pre-selected transmission functions by simply tuning a voltage.  We combine this bi-colour functionality in a nonlinear optical experiment by employing three-wave mixing in a periodically poled crystal to show how the combined effect of linear spin-orbit transformation rules and nonlinear selection rules gives rise to novel approaches for light to modulate light, and light to unravel light.  We show that the roles of the nonlinear crystal and spin-orbit device can be switched to either characterise the device with known light, or unravel unknown light with the device.
    This synergy between spin-orbit and nonlinear optics offers a novel paradigm where light manipulates and reveals its own structure across spectral domains.\\
\end{abstract}

%\setboolean{displaycopyright}{false} % Do not include copyright or licensing information in submission.

\maketitle

\section{Introduction}
Light can be tailored in various degrees of freedom, so-called structured light~\cite{forbes2021structured}, mostly with optical toolkits such as spatial light modulators \cite{yang2023review} and metasurfaces~\cite{dorrah2022tunable,yu_spinorbitlocking_2025}, giving rise to many exciting new fields and applications \cite{angelsky2020structured,wang2021generation}. Recently, the focus has shifted towards nonlinear control of structured light \cite{buono2022nonlinear, li2017nonlinear,de_ceglia_nonlinear_2024,gao_topology_imprinting_2025}, beyond just frequency conversion and towards new applications that include high-dimensional quantum \cite{sephton2023quantum,qiu2023remote}  and classical communication \cite{xu2023orthogonal},  real-time error correction~\cite{singh2024light}, holography \cite{ackermann2023polarization, liu2020nonlinear,yesharim2023direct}, novel imaging techniques \cite{qiu2018spiral,hong2020second,wang2021mid,sephton2019spatial}, and new paradigms \cite{da2022observation,abrahao2024shadow,wu2019vectorial,wu2022conformal,martin2025extreme}.  A modern trend is to mix degrees of freedom to create new forms of light, for instance, optical topologies \cite{shen2024optical,luttmann2023nonlinear}, spatiospectral \cite{kopf2023correlating} and spatiotemporal light \cite{zhan2024spatiotemporal}, as well as emulating quantum states by nonseparable states of light \cite{shen2022nonseparable,karnieli2021emulating}.

A common route to generate and analyse structured light has been to couple polarisation and orbital angular momentum (OAM), with spin-orbit optical devices playing a central role in both the creation and detection of such states \cite{bliokh_natphot_2015,cardano_natphot_2015,barboza_aop_2015,rubano2019q, hakobyan2025q,deng2020full}. These devices have traditionally operated in linear optics and at a single wavelength \cite{chen2023spin,chen2020high,wu2023optical,wang2024spin}. Yet, combining wavelength, polarisation, and OAM degrees of freedom is known to offer enhanced functionality and richer light–matter interactions \cite{kopf2023correlating,liu2024spatiotemporal,litchinitser2012structured}. Examples include pulse characterisation \cite{kopf_spectral_2021,jolly_spatio-spectral_2021}, photonic quantum information processing \cite{slussarenko_photonic_2019, flamini_photonic_2019} and cascade of spin-orbit states \cite{tang2020harmonic,tang2025sequential}, where all degrees of freedom are needed to leverage the benefits of high information density.

Here we use a spin-orbit liquid crystal device that can tailor controlled superpositions of wavelength, polarisation, and OAM states simultaneously. In this work, we probe this device for two wavelengths sequentially in a proof-of-principle experiment, which we refer to as bi-colour. This could be extended to a dual-wavelength set up such that a superposition of all three degrees of freedom can be controlled. 
The bi-colour behaviour lends itself to nonlinear creation and detection of structured light, which we demonstrate by characterising the modal properties of light in a quadratic nonlinear optical process, with access to both wavelengths while only measuring in the more easily detectable output wavelength. The device can be positioned either before or after the nonlinear stage, enabling two complementary modes of modulation---each shaping the detection of the light’s internal structure. This spin-orbit bi-colour modulation can be extended to any nonlinear process and can be tuned to the relevant wavelengths, without restriction on the number of wavelengths involved. Especially in collinear processes, it will be experimentally easy to have several wavelengths propagating through the device simultaneously. The interplay of multi-degree-of-freedom control with multi-colour spin-orbit coupling in a nonlinear context suggests novel routes for advanced manipulation and decoding of structured light beyond bi-colour spin-orbit-based imaging applications \cite{yan_optica_2015, arjmand_acsphot_2025}.

\section{Spin-orbit device and nonlinear interaction stages}
\begin{figure*}[b]
	\centering	\includegraphics[width=0.6\linewidth]{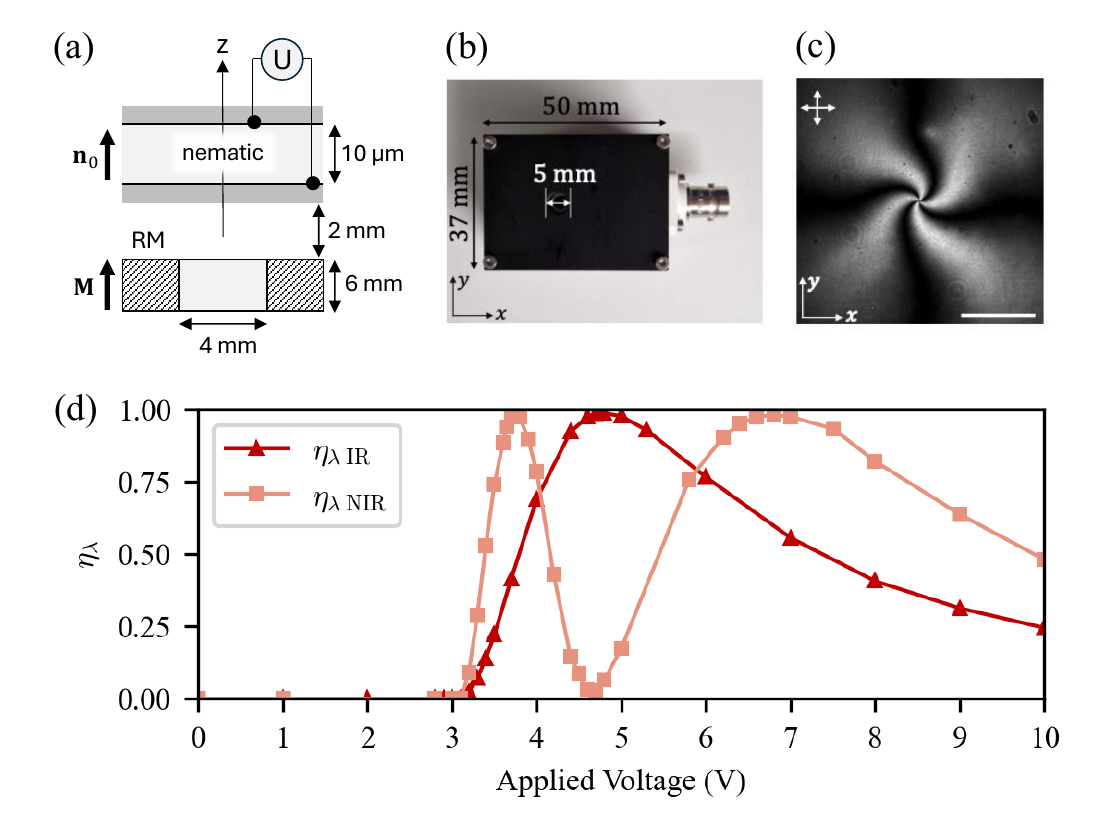}
	\caption{(a) Side-view sketch of the spin-orbit liquid crystal device made of a nematic slab with uniform orientation at rest $\bm{n_0}$ aligned with the $z$ axis, sandwiched between two glass substrates provided with transparent electrodes delivering a square waveform voltage $U$ at 100 kHz. The device is operated using the combined action of a uniform quasistatic electric field and a structured static magnetic field obtained from a ring magnet (RM) with uniform magnetisation $\bm{M}$. Not to scale. (b) Top-view image of the packaged device. (c) Observation of the topological structure of the nematic liquid crystal slab under an applied voltage of 4.70 V by optical imaging between crossed linear polarisers. Scale bar: 300~$\upmu$m. (d) Spin-to-orbital conversion efficiency efficiency, $\eta_\lambda$, as a function of applied voltage for $\lambda_{\mathrm{NIR}} = 808$~nm and $\lambda_\mathrm{IR} = 1550$~nm. Solid curves are guides for the eyes.}
	\label{fig:bumbilic_fab}
\end{figure*}

The spin-orbit device, illustrated in Figs.~\ref{fig:bumbilic_fab}(a) and \ref{fig:bumbilic_fab}(b), consists of a nematic liquid crystal (LC) cell combined with a cylindrical ring magnet (height: 6~mm, outer diameter: 8~mm, inner diameter: 4~mm) placed at a distance $d\simeq 2$~mm from the LC cell (EHC Co. Ltd., Japan) having a thickness of $L = 10 \pm 0.5~\upmu$m and is equipped with indium-tin-oxide electrodes for voltage control. The liquid crystal has an elastic relaxation time of $\approx$ 0.1 s. The actual time for switching from one state to another depends on the initial and final state as well as the temporal waveform of the envelope of the applied quasistatic voltage. Both confining glass substrates impose homeotropic anchoring, ensuring a uniform initial molecular alignment along the optical axis ($z$-direction). The cell is filled with the nematic mixture 1859A (MUT, Poland), which exhibits a birefringence of $\Delta n \simeq 0.229$ at a wavelength of $\lambda = 589$~nm and a negative dielectric anisotropy at 100~kHz frequency. 

As demonstrated in the previous work \cite{brasselet_prl_2018}, under the combined action of magnetic and electric fields, a stable macroscopic topological defect with charge $q=+1$ spontaneously forms in the nematic, as shown in Fig.~\ref{fig:bumbilic_fab} (c). The axisymmetric magnetic field seeds the peripheral reorientation of the liquid crystal molecules. This defect behaves like a space-variant waveplate: away from its core, the birefringence is nearly uniform, and by tuning the voltage one can reach the half-wave condition.  It is an effective uniaxial slab with a tunable uniform birefringent phase retardation~$\Delta$, and an azimuthally varying optical axis orientated at an angle $\psi = q\varphi$ from the $x$-axis, where $\varphi$ is the polar angle in the $(x,y)$ plane. The optical vortices with topological charges $l=\pm2$ are generated for left- and right-circularly polarised incident Gaussian beams, provided that the phase retardation satisfies $\Delta = N\pi$, with $N$ being an odd integer \cite{marrucci_optical_2006, brasselet_electrically_2011}. In general, the device performs the following transformations:

\begin{equation}
\label{eq:transformation1}
\ket{l, \sigma^{\pm}, \lambda}\longrightarrow \Big( \sqrt{1-\eta_\lambda}\ket{l, \sigma^{\pm}} + \sqrt{\eta_\lambda}\ket{l \pm 2, \sigma^{\mp}}\Big)\ket{\lambda},
\end{equation}

\noindent where $\sigma^+$ ($\sigma^-$) corresponds to the left (right)-circular polarisation states of wavelength $\lambda$, and $\eta_\lambda = \sin^2(\Delta/2)$ is the vortex conversion efficiency. We recall that $\Delta(\lambda) = 2 \pi L dn/\lambda$, where $dn$ is the effective birefringence, which not only depends on the applied voltage but also has dispersion. More detailed analyses are reported in \cite{brasselet_electrically_2011, hakobyan_hyperspectral_2024}. The electric field allows us to tune $\Delta$, hence the conversion efficiency, which enables control over the weights of the output light components. The vortex transformations also apply when a superposition of left- and right-circular polarisation is used
\begin{comment}
\begin{equation}
    \begin{split}
        \ket{l}\Big(\sqrt{\alpha}\ket{\sigma^+} + \sqrt{\beta}\ket{\sigma^-} \Big)\ket{\lambda}\longrightarrow & \Bigg( \sqrt{1-\eta_\lambda}\ket{l} \Big(\sqrt{\alpha}\ket{\sigma^+} + \sqrt{\beta}\ket{\sigma^-} \Big) + \\
        &\sqrt{\eta_\lambda}\Big(\sqrt{\alpha}\ket{l+2, \sigma^-} + \sqrt{\beta}\ket{l-2,\sigma^+} \Big) \Bigg) \ket{\lambda},
    \end{split}
    \label{eq: superposition polarisaiton transformation rule}
\end{equation}
\end{comment}
\begin{equation}
    \begin{split}
        \ket{l}\Big(\sqrt{\alpha}\ket{\sigma^+} + \sqrt{\beta}\ket{\sigma^-} \Big)\ket{\lambda}\longrightarrow & \Bigg( \sqrt{1-\eta_\lambda} \Big(\sqrt{\alpha}\ket{l,\sigma^+} + \sqrt{\beta}\ket{l,\sigma^-} \Big) + \\
        &\sqrt{\eta_\lambda}\Big(\sqrt{\alpha}\ket{l+2, \sigma^-} + \sqrt{\beta}\ket{l-2,\sigma^+} \Big) \Bigg) \ket{\lambda},
    \end{split}
    \label{eq: superposition polarisaiton transformation rule}
\end{equation}
where $\alpha$ and $\beta$ represent the coefficients of the circularly polarised states.
This device, therefore, functions as an electrically tunable q-plate but distinguishes itself by its self-engineered liquid crystal topological defect, in contrast to the mainstream technological approaches consisting of artificially patterned orientation structures. This comes with structural assets, such as a preserved structural axisymmetry down to the core of the topological defect, as discussed in the previous work \cite{brasselet_prl_2018}.  This enables both high structural quality and tuneable voltage features, in contrast to conventional q-plates, whether they are manufactured \cite{rubano2019q} or obtained via previous self-engineering strategies \cite{barboza_aop_2015}. In particular, this allows manipulating more than one wavelength simultaneously, as already exploited in the context of optical imaging \cite{yan_optica_2015, arjmand_acsphot_2025}. For two input beams of wavelengths $\lambda_1$ and $\lambda_2$, the output becomes

\begin{equation}
    \begin{split}
       \ket{l_1,\sigma^\pm,\lambda_1}\otimes \ket{l_2,\sigma^\pm,\lambda_2} \longrightarrow & \Big(\sqrt{1-\eta_{\lambda_1}}\ket{l_1, \sigma^\pm} + \sqrt{\eta_{\lambda_1}}\ket{l_1\pm2, \sigma^\mp}\Big)\ket{\lambda_1} \\
       \otimes & \Big(\sqrt{1-\eta_{\lambda_2}}\ket{l_2, \sigma^\pm} + \sqrt{\eta_{\lambda_2}}\ket{l_2\pm2, \sigma^\mp}\Big)\ket{\lambda_2},   
    \end{split}
    \label{sorule}
\end{equation}

Note that each respective efficiency $\eta_\lambda$ depends uniquely only on the applied electric field, and $\Delta$ is wavelength-dependent. Hence, bimodal behaviour, i.e. two distinct topological charge output states, can be achieved by careful selection of the input states and the applied voltage, allowing spin-orbit coupling for individual colour modes simultaneously. 
In our study, the thickness and the type of liquid crystal were chosen to provide ``full/zero" conversion for infrared (IR) / near-infrared (NIR) wavelengths for a given voltage (4.7 V). The distance between the magnet and the cell corresponds to an experimentally optimised configuration. A detailed discussion of the dependence of this parameter can be found in \cite{brasselet_prl_2018}.

Fig. \ref{fig:bumbilic_fab}(d) shows the device's efficiency, $\eta_{\lambda}$, as a function of applied voltage for the two wavelengths of interest in this study, namely for $\lambda_{\mathrm{NIR}} = 808$~nm and $\lambda_\mathrm{IR} = 1550$~nm. To characterise $\eta_{\lambda}$, we measured the output power fraction, $P_{\lambda}^{\sigma^-}$, for input polarisation state $\sigma^+$ following $\eta_{\lambda} = P_{\lambda}^{\sigma^-}/(P_{\lambda}^{\sigma^+} + P_{\lambda}^{\sigma^-})$.  The results for wavelengths $\lambda_{\mathrm{NIR}} = 808$~nm and $\lambda_\mathrm{IR} = 1550$~nm demonstrate the tuneability of the conversion efficiency with applied voltage. In particular, an applied voltage of 4.70 V yields a maximal conversion efficiency of $\eta_{\lambda_\mathrm{IR}} =  0.98$, whilst $\eta_{\lambda_\mathrm{NIR}} = 0.03$ is minimised. These respective efficiencies allow for the creation of the bimodal state $\ket{2, \sigma^-, \lambda_{\rm IR}} \otimes \ket{0, \sigma^+, \lambda_{\rm NIR}}$. For $\eta_{\lambda_\mathrm{NIR}}$, we observe maximal efficiencies of 0.99 and 0.98 at 3.75 V and 6.80 V applied voltage, respectively. At these applied voltages, $\eta_{\lambda_\mathrm{IR}}$ is $\simeq$ 0.4. Simultaneous efficiency optimisation for both wavelengths can be achieved at 4.04 V, with $\eta_{\lambda_\mathrm{NIR}} = \eta_{\lambda_\mathrm{IR}} = 0.72$, and at 5.89 V, with $\eta_{\lambda_\mathrm{NIR}} = \eta_{\lambda_\mathrm{IR}} = 0.79$.

We now describe how this spin-orbit device is integrated with a nonlinear optical process. Anticipating the experiment to follow, we consider the case of a type-0~$\chi^{(2)}$ crystal where the only requirement is polarisation matching, $\mathbf{e}+\mathbf{e}\to \mathbf{e}$. Here $\mathbf{e}$ is a unit vector that denotes the polarisation state \cite{manjooran2012phase}. In this configuration, difference-frequency generation (DFG) occurs between two copolarised input fields at frequencies $\omega_1$ and $\omega_2$, yielding a new frequency $\omega_\mathrm{DFG}=\omega_1-\omega_2$.  In the context of structured light, the output electric field can be written as $\mathbf{E}_\mathrm{DFG} \propto \mathbf{E}_1 \times \mathbf{E}_2^*$, where the complex conjugate introduces a phase difference that encodes the structure of both input fields.
In bra-ket notation, the input state can be represented as a tensor product $ \ket{l_1,\sigma^{\pm},\omega_1} \otimes \ket{l_2,\sigma^{\pm},\omega_2}$, and the resulting DFG output becomes
\begin{equation}
\mathbf{E}_\text{DFG} \equiv \ket{l_1-l_2,\,\mathbf{e},\, \omega_1 - \omega_2}
\end{equation}
where the output polarisation  $\mathbf{e}$ aligns with the optical axis, which is fixed by the orientation of the anisotropic nonlinear crystal. When combined with the transformation rules of Eq.~(\ref{sorule}), it becomes clear that the device enables independent modulation of both wavelengths involved in the nonlinear interaction. The interplay between linear spin-orbit transformations and nonlinear selection rules thus creates new opportunities for both structuring light and probing structured light in nonlinear optical processes. We now turn to the experimental realisation of this concept.

\section{Experimental set-up}

\begin{figure*}[b]
	\centering	\includegraphics[width=0.8\linewidth]{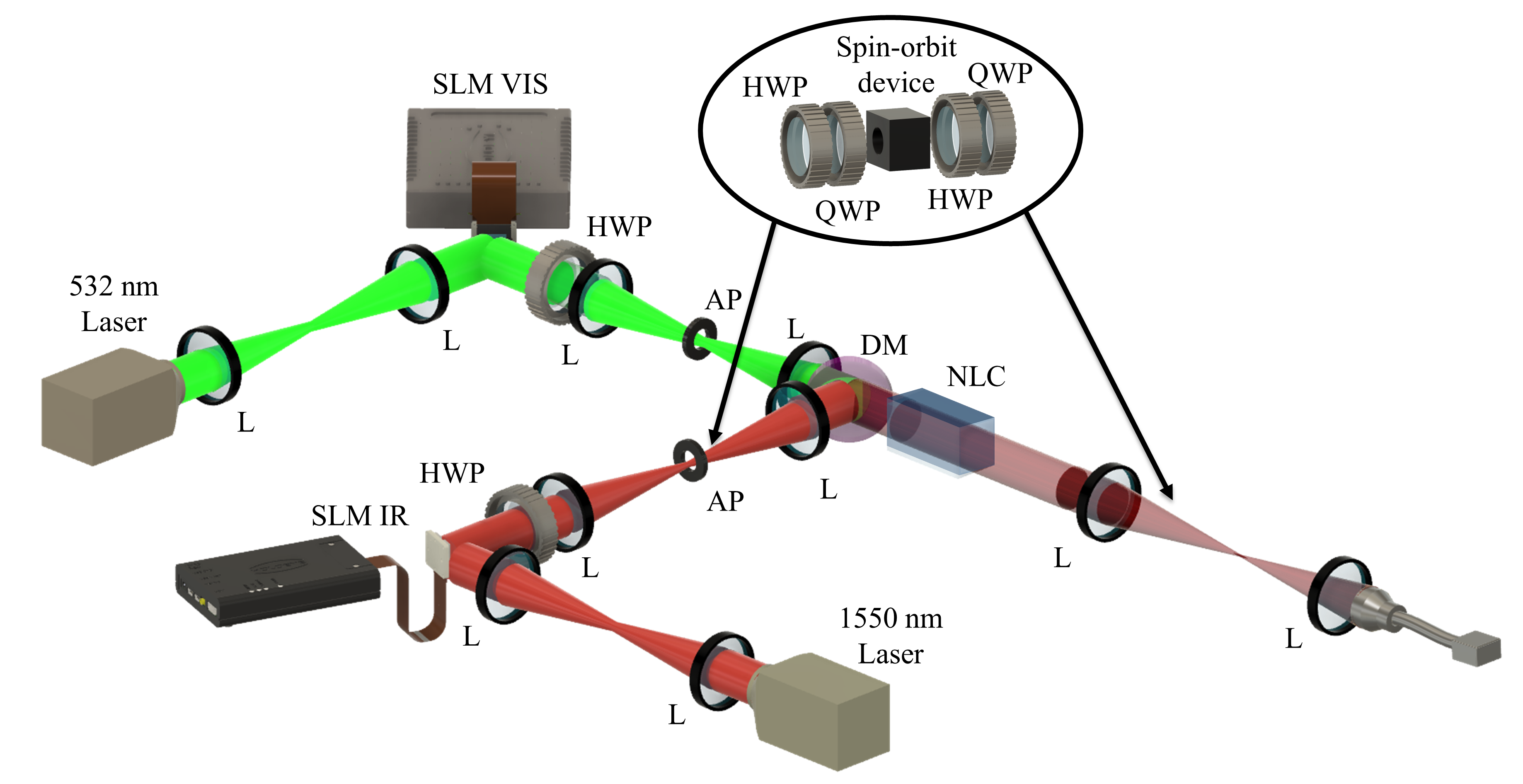}
	\caption{Schematic of the experimental setup for nonlinear spatial mode detection with a spin-orbit device. A 532 nm (VIS) beam and a 1550 nm (IR) beam are expanded and collimated using a 4$f$-lens system before being modulated by spatial light modulators ($\mathrm{SLM_\mathrm{VIS}}$ and $\mathrm{SLM_\mathrm{IR}}$). Half-wave plates (HWP) adjust the polarisation before the beams are combined in a nonlinear crystal (NLC) via a dichroic mirror (DM). The DFG process in the NLC produces an 808 nm (NIR) beam, whose spatial mode structure is analysed using a single-mode fibre (SMF) and an avalanche photodiode (APD). The inset shows that the spin-orbit device is placed either in the NIR or IR arm. HWPs and quarter-wave plates (QWP) control the ingoing and outgoing polarisation.}
	\label{fig:expsetup2}
\end{figure*}

The spin-orbit device was placed into the experimental setup shown in Fig.~\ref{fig:expsetup2}. A type-0 periodically poled KTP nonlinear crystal (NLC) was operated in DFG mode to generate an output beam of wavelength $\lambda_\text{NIR} = 808$ nm for input beams with wavelengths $\lambda_\mathrm{VIS}$ = 532 nm and  $\lambda_\mathrm{IR} = 1550$ nm.  In this work, we target vortex conversion for wavelengths $\lambda_{\mathrm{NIR}}$ and $\lambda_\mathrm{IR}$ for which the device can be tuned electrically to operate on either a single wavelength or both simultaneously. The complex amplitude of the IR (0.7 W) and VIS (0.6 W) beams was controlled using two spatial light modulators (SLMs), SLM IR (Holoeye PLUTO) and SLM VIS (Holoeye  GAEA 2), respectively. The VIS beam was collimated and expanded using a 4$f$ system to overfill SLM VIS, while the Gaussian IR beam was simply collimated. The output from the two SLMs was relayed onto the front face of the NLC using telescopic 4$f$ lens configurations. Half-wave plates (HWP) were used to align the polarisations for optimal phase matching. A dichroic mirror (DM) was used to overlap the two beams collinearly within the NLC. Finally, the DFG output was imaged into a single-mode fibre (SMF) and detected with an avalanche photodiode (APD) to perform an optical inner product measurement for modal decomposition \cite{pinnell2020modal}. In all experiments, a 100 kHz on/off voltage was applied to the spin-orbit device, and all voltages are quoted as root mean square values.

As shown in Fig.~\ref{fig:expsetup2}, the main experiment featured two configurations for the use of the spin-orbit device: in the first, the device was used \textit{before} the crystal to modulate the $\lambda_\mathrm{IR} = 1550$ nm, thereby altering the nonlinear interaction of the two input beams. In this configuration, the device acts as a shaping element, structuring the IR input to have OAM of $l_\text{IR}$.  To test this efficacy, we use the VIS beam as a modal detector, exploiting the conjugate nature of the DFG process.  Since the DFG output is coupled into a single-mode fibre (admitting only a Gaussian beam with $l_\text{NIR} = 0$), the selection rule for detection requires $l_\text{IR} = l_\text{VIS}$.

In the second configuration, the spin-orbit device was placed \textit{after} the nonlinear crystal, and used to modulate the $\lambda_\text{NIR}$.  
Here the spin-orbit device (SOD) modifies the OAM of the DFG beam by adding an amount $l_\text{SOD}$ to the output from the crystal. Thus, the final OAM becomes $l_\text{NIR} + $ $l_\text{SOD}$, where $l_\text{NIR}$ $= l_\text{VIS} - l_\text{IR}$ as a result of the nonlinear selection rule.
Again, the output is coupled into a single-mode fibre to assess the performance via modal decomposition. These two configurations, combined with modal analysis as a performance metric, demonstrate that the device can both shape and detect structured light at two wavelengths, opening new avenues for nonlinear control and probing of complex light fields.

\section{Results}
\begin{figure}[h]
	\centering
	\includegraphics{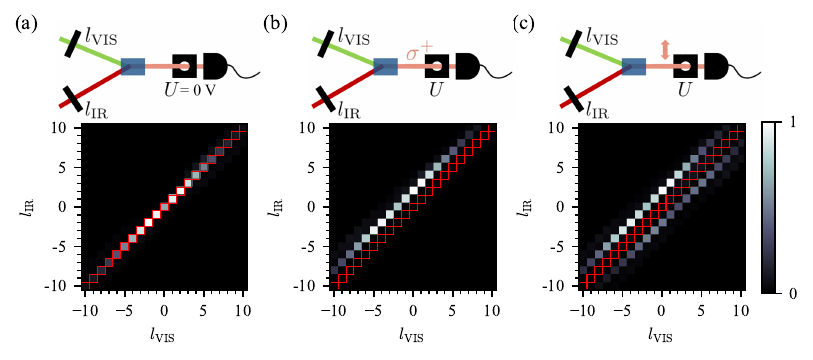}
	\caption{Modal decomposition with the spin-orbit device in the 808 nm arm of the nonlinear system.   After the crystal, $l_\mathrm{NIR} = l_\mathrm{VIS} - l_\mathrm{IR}$. When $l_\mathrm{NIR}= 0$ after the device,  maximal counts are detected. (a) Without an applied voltage, the spin-orbit coupling leaves the $l_\mathrm{NIR}$ unaltered. When the optimal NIR voltage of 3.75~V is applied, we observe the device adds a topological charge of (b) 2 for incident left circular light and (c) an equal superposition of 2 and -2 for incident linear light. We infer these OAM conversions from the diagonal shifts of maximal detection events in the modal decomposition. The red square outlines indicate when $l_\mathrm{VIS} = l_\mathrm{IR}$. The insets show the schematic set ups for recording this data.}
	\label{fig: modal_decomp_808 nm}
\end{figure}

\begin{figure}[t]
	\centering
	\includegraphics{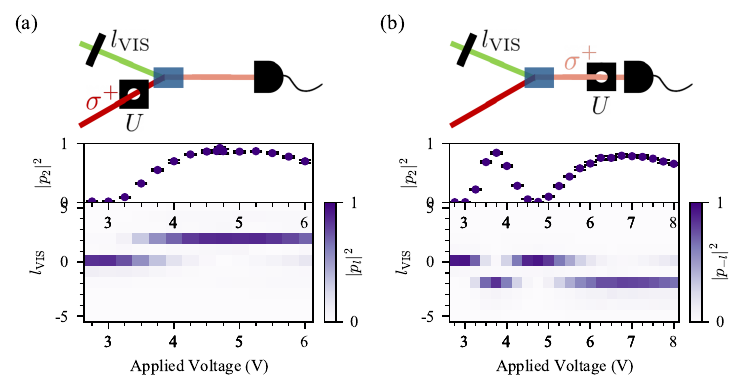}
	\caption{Characterisation of the spin-orbit device as a function of applied voltage for wavelengths (a) 1550 nm and (b) 808 nm. Modal decomposition is illustrated in the bottom plots, showing the mean detection counts of 11 repeats for different $\mathrm{LG}_{p=0, \ l}$ modes with $l$ ranging from -5 to 5 prepared on SLM VIS. (a) Detection counts are maximised when $l_\mathrm{IR}$ after the device equals $l_\mathrm{VIS}$. We observe that for a left-circularly polarised input beam, a topological charge of 2 is added by the device. (b) Here $l_\mathrm{NIR}=l_\mathrm{VIS}$, and detection counts are maximised when there is no OAM remaining in the NIR beam after the device. Again, we observe that for a left-circularly polarised input beam, a topological charge of 2 is added by the device. The top plots illustrate the modal purity of the $l=2$ component after the spin-orbit device as a function of voltage. Highest modal purities were found to be (a) 0.92 $\pm$ 0.02 at 4.70~V and (b)~0.84~$\pm$~0.02 at 3.75~V. ``Full/zero" conversion for IR/NIR is achieved at $\approx$ 4.70 V. The insets show the schematic setups for recording this~data.}
	\label{fig:voltage_characterisation}
\end{figure}

\begin{figure*}[t]
    \includegraphics{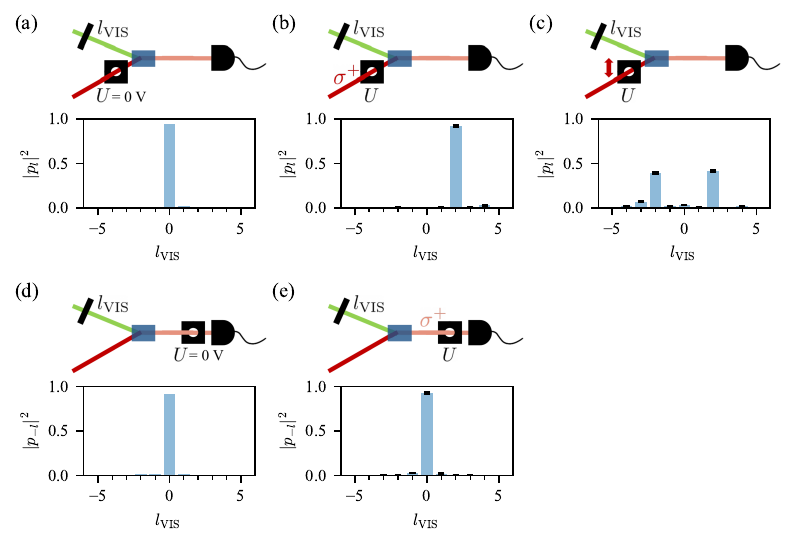}
	\caption{Modal decomposition at ``full/zero" conversion voltage in the 1550 nm (a, b, c) and 808 nm (d, e) arm of the nonlinear system. 
    Error bars show the standard deviation of 11 repeats for a 1 s measurement time. Without an applied voltage, the spin-orbit coupling leaves the (a) $l_\mathrm{ IR} = 0$ and (d) $l_\mathrm{NIR} = 0$ OAM values unaltered. When the optimal ``full/zero" conversion voltage of 4.70~V is applied, we observe the spin-orbit device converts $l_\mathrm{IR}$ = 0 to (b) $l_\mathrm{IR}$ = 2 for incident left circularly polarised light and (c) an equal superposition of $l_\mathrm{IR} = \{-2, \, 2\}$ for linearly polarised incident light. (e) The NIR beam remains unaffected at a voltage set to $\approx$ 4.7 V, corresponding to the flat region of the curve in Fig.~ \ref{fig:voltage_characterisation} (b). The insets show the schematic setups for recording this data.}
	\label{fig: modal_decomp_1550nm}
\end{figure*}

We begin by validating the known OAM nonlinear rules [Fig. ~\ref{fig: modal_decomp_808 nm} (a)] and known linear rules pertaining to q-plates, shown in Figs. ~\ref{fig: modal_decomp_808 nm} (b) and (c), by showing a modal decomposition with the spin-orbit device in the 808 nm arm of the nonlinear system. We then characterise the modal performance of the device as a function of voltage for both 1550 nm (Fig. \ref{fig:voltage_characterisation} (a)) and 808 nm (Fig. \ref{fig:voltage_characterisation} (b)), highlighting the capability of ``full/zero" conversion for the IR/NIR wavelengths (Fig. \ref{fig: modal_decomp_1550nm}). \\

Fig.~\ref{fig: modal_decomp_808 nm} shows the full modal spectrum when the two input beams, VIS and IR, were shaped by SLMs and the DFG was shaped by the spin-orbit device. A schematic of the configuration in the inset is shown to aid the reader. The SLMs were programmed to produce all combinations of $l_\mathrm{VIS}$, $ l_\mathrm{IR} \in \{-10,10\}$ in the Laguerre-Gaussian basis by complex amplitude modulation and with no radial mode, i.e., $p = 0$ for all modes. When the device was turned off, Fig.~\ref{fig: modal_decomp_808 nm} (a), the result is a cross-talk matrix with only a diagonal signal, highlighted in red for visualisation. The cross-talk matrix indicates good input mode quality from the SLMs and validates known OAM behaviour in three-wave mixing. When the device was turned on, the functionality depended on the voltage, incoming polarisation and wavelength.  We modulated the NIR beam at the optimal voltage for full NIR conversion for two incoming polarisations, left-circular and linear, with the results shown in Figs.~\ref{fig: modal_decomp_808 nm} (b) and (c).  We observe an addition of topological charge 2 for left circularly polarised light, evident by the shift of the diagonal counts by +2, in agreement with the transformation rules in Eq.~\eqref{eq:transformation1}.  For incident linearly polarised light, the spin-orbit device both adds and subtracts a topological charge of 2 to the input mode, resulting in a dual offset in the diagonal, in agreement with the transformation described in Eq.~\eqref{eq: superposition polarisaiton transformation rule}. The original state is shown in red for visualisation. The slightly lower counts for the downshifted curve might be due to alignment or small polarisation~mismatch.\\

Next, we characterised the modal performance of the device as a function of voltage, with the results shown in Fig.~\ref{fig:voltage_characterisation}. We define modal purity in the OAM basis as the fractional power $|p_l|^2$ in the expansion $\sum p_l \exp(il\phi)$, where $\phi$ is the azimuthal angle, with the weights normalised so that $\sum |p_l|^2 = 1$.  In the first experiment, the device was placed in the left-circularly polarised IR beam with SLM IR switched off, so that all the IR modulation was due to the spin-orbit device. 
In the second experiment, we kept SLM IR switched off and moved the device to after the nonlinear crystal to modulate the DFG beam.  In both cases, SLM VIS was programmed to produce $l_\text{VIS} \in \{-5,5\}$ in the Laguerre-Gaussian basis by complex amplitude modulation and with no radial mode, i.e., $p = 0$ for all modes.  The resulting detection into the single mode fibre ensured that the device's response could be uniquely determined by this modal analysis.  The two set ups are schematically portrayed in the insets of Figs.~\ref{fig:voltage_characterisation}~(a) and (b), respectively. The top panel in each plot shows the modal power in the expected $l = 2$ mode, $|p_2|^2$, with the bottom panels showing the full modal data, $|p_l|^2$, both as a function of voltage. In Fig.~\ref{fig:voltage_characterisation} (a), since the device was placed before the crystal, detection was only possible when the OAM of the device and $l_\text{VIS}$ match, here only when $l_\text{VIS} = 0$ or 2, tuned by the efficiency of the device.  The highest modal purities were found to be $|p_2|^2 = 0.92 \pm 0.02$ at 4.70~V.  Similar performance was found when the device modulated the DFG output beam, with the results shown in Fig.~\ref{fig:voltage_characterisation} (b).  In this case, detection is only possible when the OAM of the device and $l_\text{VIS}$ are conjugated, here only when $l_\text{VIS} = 0$ or $-2$, tuned by the efficiency of the device.  The highest modal purity was found to be $|p_2|^2 = 0.84 \pm 0.02$ at 3.75 V. The slightly lower purity when placing the device after the crystal may be due to aberrations and distortions introduced by the crystal itself, caused for example by thermal effects or minor imperfections in the crystal. Another potential reason could be the discrete nature of the voltages used in the experiment.  Small changes in voltage result in small changes to the purity, and, thus, a higher modal purity may have been found between the voltage settings used.\\

We now highlight the performance for ``full/zero" conversion for IR/NIR, which is possible at $\approx$ 4.70 V, as presented in Fig.~\ref{fig: modal_decomp_1550nm}.  Once again, we show a schematic of the configuration in the inset to aid the reader. With the device switched off, shown in Fig.~\ref{fig: modal_decomp_1550nm} (a) [(b)], only $l_\text{IR} = 0$ [$l_\text{NIR} = 0$] was detected.  At the ``full/zero" conversion voltage, changing the input polarisation altered the final state for IR, clearly showing $l_\text{IR} = 2$ for left-circularly polarised light, and an equal weighting of $l_\text{IR} = \pm 2$ for linearly polarised light, as seen in Figs.~\ref{fig: modal_decomp_1550nm} (b) and (c), respectively.  However, the NIR is unaffected when this voltage is applied to the device, as shown in Fig.~\ref{fig: modal_decomp_1550nm} (e). This demonstrates the proof-of-principle that when the spin-orbit-coupling device is subjected to both NIR and IR wavelengths, it can leave the former unaffected, while fully converting the latter, enabling the creation of a bimodal state.

\newpage

\section{Conclusion}
We exploited a wavelength-tunable spin-orbit device that flips the helicity of circularly polarised input light and changes its OAM state by amounts of two units, in a nonlinear optics framework. This is achieved by a stable topological defect in a nematic slab that results from the combined action of magnetic and electric fields.
We show this device can perform spin-orbit coupling for 1550 nm with efficiency $\eta_{\lambda_\mathrm{IR}} =  0.98$ but leaves 808 nm unaffected ($\eta_{\lambda_\mathrm{NIR}} =  0.03$). We have demonstrated this for the wavelengths individually. Placing the device in a dual wavelength set-up would enable the generation of a bimodal state that is wavelength-dependent. Additionally, both wavelengths can be subjected to spin-orbit coupling simultaneously. 
We reported an efficiency~$\eta_\lambda (U)$ look-up table that reports how their respective efficiencies depend on the applied electric field. 
For 808 nm, we can achieve maximum efficiency $\eta_{\lambda_\mathrm{NIR}} =  0.99$ ($\eta_{\lambda_\mathrm{IR}} \simeq 0.4$). From the individual efficiency look-up tables, we can identify that the maximum efficiency for a simultaneous bi-colour spin-orbit coupling would be around $\eta_{\lambda_\mathrm{NIR}} = \eta_{\lambda_\mathrm{IR}} = 0.79$. 

Summarising, we characterised how a spin-orbit device can affect the modal purity of an input OAM spectrum for $\lambda_\mathrm{IR}$ and $\lambda_\mathrm{NIR}$ in a nonlinear DFG process such that modal decomposition measurements were performed exclusively in the more easily accessible NIR spectral range. The modal purity of generated OAM mode as a function of the applied voltage exhibited a response mirroring $\eta_\lambda$, which quantifies the power fraction of an incident circularly polarised component that is processed by the spin-orbit interaction. High-purity OAM conversions by the spin-orbit device at the voltage values corresponding to maximal efficiency were observed for both circularly and linearly polarised input light, in agreement with theoretical predictions. These results highlight that a wavelength-tunable spin-orbit device opens the door for a wide range of applications in structured light and quantum information, such as light-matter interactions, pulse characterisation, quantum imaging, and maximising information density. In particular, the superposition coefficients of hybrid states $\ket{l, \mathbf e, \lambda}$ can be precisely controlled and thus, combining dual-wavelength bimodal states with nonlinear three-wave mixing may broaden the accessible spectral range for modal detection of structured light \cite{sanchez2025nonlinear}.  Furthermore, output bimodal states can be generated with extremely high efficiencies.

\section*{Acknowledgements}
The authors would like to thank the CNRS-Wits collaboration fund.

\section*{References}
% Bibliography
\bibliography{references}

\end{document}